\documentclass[onecolumn,preprintnumbers,nofootinbib,amsmath,amssymb,superscriptaddress,11pt,a4paper]{article}

\usepackage{amsmath}
\usepackage{graphics}
\usepackage{amsfonts}
\usepackage{amssymb}%
\usepackage{enumitem}
\usepackage{geometry}
\usepackage{graphicx}
\usepackage{dcolumn}
\usepackage{setspace}
\providecommand{\U}[1]{\protect\rule{.1in}{.1in}}

\newtheorem{Ass'}{Assumption'}

\newcommand{\be}{\begin{equation}}
\newcommand{\ee}{\end{equation}}

\newcommand{\hammer}{\rm Hammer}
\newcommand{\winp}{\rm wp}

\onehalfspace
\begin{document}
\title{Elementary econometric and strategic analysis of curling matches}

\author{John Fry\footnote{Centre for Mathematical Sciences, School of Natural Sciences, University of Hull, Hull, HU6 7RX, UK.\newline
Email: J.M.Fry@hull.ac.uk},$\quad$ Mark Austin\footnote{Centre for Mathematical Sciences, School of Natural Sciences, University of Hull, Hull, HU6 7RX, UK.\newline
Email: m.austin-2020@hull.ac.uk}$\quad$ and$\quad$ Silvio Fanzon\footnote{Centre for Mathematical Sciences, School of Natural Sciences, University of Hull, Hull, HU6 7RX, UK.\newline
Email: S.Fanzon@hull.ac.uk}}

 \date{October 2024} \maketitle \setcounter{tocdepth}{2}
\begin{abstract}
We develop a Markov model of curling matches, parametrised by the probability of winning an end and the probability distribution of scoring ends. In practical applications, these end-winning probabilities can be estimated econometrically, and are shown to depend on which team holds the hammer, as well as the offensive and defensive strengths of the respective teams. Using a maximum entropy argument, based on the idea of characteristic scoring patterns in curling, we predict that the points distribution of scoring ends should follow a constrained geometric distribution. We provide analytical results detailing when it is optimal to blank the end in preference to scoring one point and losing possession of the hammer. Statistical and simulation analysis of international curling matches is also performed.
\end{abstract}






\section{Introduction}

There is a sustained interest in sports analytics (Baker et al., 2022, Singh et al., 2023; Scarf et al., 2022). Curling, in particular, has garnered significant academic attention (Lawson and Rave, 2020; Willoughby and Kostuk, 2004; 2005), with a focus on statistical and strategic analyses (Kostuk et al., 2001; Brenzel et al., 2019). For example, Kostuk et al. (2001) discuss representative scoring patterns in curling,  while Brenzel et al. (2019) offer an insightful analysis of whether winning an end by a single point is advantageous, given that it involves ceding strategic control of the hammer. Abstracting from key features of the game, we develop a Markov model for curling matches parameterised by the end-winning probabilities and the points distribution of scoring ends. This model addresses strategic and econometric questions and offers insights into the theoretical aspects of the game's dynamics.

In the curling match, as outlined in Section~2, determining whether winning ends with a single stone are advantageous poses a key strategic dilemma in the sport. For instance, Brenzel et al. (2019) offer an elegant computational solution to this problem. However, this solution may be challenging to implement in practice. In this paper, we introduce a simpler analytical approach, taking advantage of the well-established fact that hammer possession in a curling match exhibits a Markovian structure. Specifically, we define a Markov chain model that characterises hammer possession and yields a solvable equilibrium distribution. This feature allows us to provide an approximate analytical solution to the curling match problem. Our solution explicitly outlines the optimal strategy in terms of the relative strength of the two teams, the precise numerical benefit associated with possession of the hammer, and the expected value of a scoring end. These strategic factors add a further dimension to the econometric analysis of the model.

The proposed model, specifically designed for empirical applications, consists of two independent components. The first component is the distribution of points in scoring ends, while the second component is a set of end-winning probabilities that depend on the teams involved and on which team holds the hammer advantage. We provide an elegant solution to the first problem using the method of maximum entropy (Visser, 2013). The second problem is addressed by fitting generalised linear models to historical data, which adds sporting insights to a purely statistical analysis (Fry et al., 2021). Through this approach, we measure teams' offensive and defensive strengths, gaining new insights into the interaction between offensive and defensive strategies in elite-level curling. Additionally, the model allows us to quantify the effects of holding the hammer, which is analogous to the home-field advantage in other sports (Boudreaux et al., 2017; Ehrlich and Potter, 2023).

The layout of this paper is as follows. Section 2 provides an overview of the game of curling. A Markov model for curling matches is then developed in Section 3. The model is deliberately structured to enable strategic analyses (Section 4) and econometric analyses (Section 5) to be performed. Section 6 concludes and discusses the opportunities for further research.

\section{Overview of the game of curling}
Curling is a sport that is played on ice between two teams. The teams take it in turns to slide stones towards a target, known as the House. Traditional teams are made up of four players, either
all men or all women. The four players are referred to as Lead, Second, Third and Fourth (usually the position occupied by the Skip). A mixed event is also present at Olympic level which is only played
as mixed doubles, i.e. one male and one female. This event has slightly different rules to that of the four-person game, and will not be analysed in this paper.

During international competitions a game of curling is played over ten ends, where an end is made up of 16 stones: 8 delivered by each team. If the two teams remain level after 10 ends, the teams play to sudden death, meaning that the next score wins. The 8 stones, two by each player, are delivered with the aim of outscoring the opposition. The team that has the last stone in an end has what is called the hammer -- potentially quite a significant strategic advantage (Brenzel et al., 2019). Holding the hammer is equivalent to the home-field advantage in conventional sports (Boudreaux et al., 2017; Ehrlich and Potter, 2023) and can be quantified using a generalised linear modelling approach in Section 5. The score at the end of each end is calculated by the number of consecutive stones that are in the house which are closer to the centre, known as the button, than any of the opposition's stone. As such, only one team can score points in an end. Further, the maximum number of points that can be scored in an end is 8. The team that scores is considered to have won the end. If no stones are in the house at the end of the end, then no points are scored, resulting this in a blank end. If a team scores points in an end, then the hammer passes onto the opposition. Instead if the
end is blank, the team that has the hammer retains the hammer.
The advantage of having the hammer in the first end is known as Last Stone First End advantage and is
decided by a draw-shot challenge. This is a trial of skill at the start of the match, markedly different from e.g. the coin toss in soccer, and  entails two players from each team throwing one
stone each towards the house. The combined distance from the button is calculated and the team with the lowest combined distance from the button will get the Last Stone First End advantage.

\section{The model}

In this section we construct a statistical model for the curling match, which abstracts from some of the sport's key features -- namely, its low scoring nature, and the strategic benefits associated with holding the hammer. The format of the model is explicitly chosen with econometric applications in mind. See Section 5.

Suppose that with probability $p_X$ the end is won by Team $X$, and with probability $p_Y$ the end is won by Team $Y$. Note that in general $p_X + p_Y \neq 1$, since the end can be blank. If a team wins an end, the number of points awarded is given by a random variable $Z$ (see below). This construction matches the relatively low scoring of curling and imagines that, in practice, only relatively few scoring patterns can occur in elite curling (see e.g. Kostuk et al., 2001). When a team holds the hammer, their chances of winning the end improve using a logistic model. Specifically, the conditional probability of Team $X$ winning with the hammer, denoted $p_{X, \hammer}$, is given by the logistic equation:
\begin{eqnarray}
    \log\left(\frac{p_{X, \hammer}}{1-p_{X, \hammer}}\right)=\log\left(\frac{p_X}{1-p_X}\right)+\beta;\qquad p_{X, \hammer}=\frac{e^{\beta}p_X}{1-p_X+e^{\beta}p_X}\,,\label{eq1}
\end{eqnarray}
where $\beta > 0$ is a fixed parameter. The conditional probability of Team $Y$ winning with the hammer, $p_{Y, \hammer}$, follows the same formula as \eqref{eq1}, with $p_X$ and $p_{X, \hammer}$ replaced by $p_Y$ and $p_{Y, \hammer}$. The logistic form in \eqref{eq1} is common to applied statistical modelling (Bingham and Fry, 2010).

Possession of the hammer then proceeds according to a time-homogeneous Markov chain $P$, under the assumption that teams retain the hammer if they do not win the end:
\begin{equation}
    P= \left(\begin{array}{ll}1- p_{X,\hammer} 
         &  p_{X,\hammer}  \\
          p_{Y,\hammer} & 1 - p_{Y,\hammer} 
    \end{array}\right) =
    \left(\begin{array}{ll}1-\frac{e^{\beta}p_X}{1-p_X+e^{\beta}p_X}
         & \frac{e^{\beta}p_X}{1-p_X+e^{\beta}p_X} \\
         \frac{e^{\beta}p_Y}{1-p_Y+e^{\beta}p_Y}& 1-\frac{e^{\beta}p_Y}{1-p_Y+e^{\beta}p_Y} 
    \end{array}\right).\label{eq2}
\end{equation}
Using the eigen equation $\pi=\pi{P}$ (Grimmett and Stirzaker, 2020) the Markov Chain in (\ref{eq2}) has stationary distribution
\begin{eqnarray}
    \pi=\left(\frac{p_Y}{p_X+p_Y}, \frac{p_X}{p_X+p_Y}\right).\label{eq3}
\end{eqnarray}
The first and second components of $\pi$ represent the long-term probabilities that Team $X$ and Team $Y$, respectively, hold the hammer in a given end.

Next, consider the points-scoring distribution $Z$ for a non-blank end. Following a maximum entropy approach, as in (Visser, 2013), we model the system based on the idea that, over time, its statistical behavior converges to a maximum entropy configuration (Bishop, 2006). For curling scoring patterns, assume the average score in a scoring end is constant, reflecting typical scoring patterns in matches (Kostuk et al., 2001):
\begin{eqnarray}
    E[Z]=\sum_{n=1}^8np_n=\mu \,,\label{eq4}
\end{eqnarray}
where $p_n:=Pr(Z=n)$.
The value of 8 in the summation in equation \eqref{eq4} comes from the 8 curling stones each team has. Since only one team can score (see Section 2), the random variable $Z$ ranges from 1 to 8. To maximise the entropy, we maximise the function
\begin{eqnarray}
J(p_1,\ldots,p_n) = -\sum_{n=1}^8 p_n\ln(p_n)-\eta \left(\sum_{n=1}^8 p_n-1 \right) + \alpha\left(\sum_{n=1}^8 n p_n -\mu \right).\label{eq5}
\end{eqnarray}
The parameter $\eta$ in equation (\ref{eq5}) is a Lagrange multiplier ensuring that the normalisation condition $\sum_{n=1}^8p_n=1$ holds. Equation (\ref{eq5}) also shows that the parameter $\alpha<0$ penalises ``unphysical" solutions that do not satisfy the constraint (\ref{eq4}). Extremising (\ref{eq5}) gives
\begin{eqnarray}
    -\ln(p_n)-1-\eta+\alpha{n}=0;\quad \ln(p_n)=\alpha{n}+C ; \quad p_n=Ae^{\alpha n} = A \theta^n\,.\label{eq6}
\end{eqnarray}
Equation (\ref{eq6}) and the constraint $\sum_{n=1}^8 p_n = 1$ give rise to a constrained geometric distribution satisfying
\begin{equation}
	\label{eq7}
\begin{aligned}
    p_n & = \frac{\theta^n}{\theta+\theta^2+\theta^3+\theta^4+\theta^5+\theta^6+\theta^7+\theta^8}\qquad  (n=1, \ldots, 8)\\
    E[Z] & =  \frac{\theta+2\theta^2+3\theta^3+4\theta^4+5\theta^5+6\theta^6+7\theta^7+8\theta^8}{\theta+\theta^2+\theta^3+\theta^4+\theta^5+\theta^6+\theta^7+\theta^8}=\mu.
\end{aligned}
\end{equation}
Notice that $\theta = e^{\alpha} < 1$ because $\alpha < 0$.
Therefore the distribution in (\ref{eq7}) has the realistic physical property that ends with higher scores are progressively less likely to occur (Brenzel et al., 2019).

\subsection{Expected scores and match outcome probabilities.}

Motivated by previous sports analytics applications (see e.g. Fry et al., 2021), it is interesting to examine expected scores and match outcome probabilities. In this respect, some analytical formulae for expected scores are possible, given the equilibrium statistical distribution of the hammer in equation (\ref{eq3}). However, complications surrounding hammer possession mean that analytical formulae for match outcome probabilities are difficult to obtain, and analysis may reduce to Monte Carlo simulation.  

Let $q_i$ represent the probability that Team $X$ holds the hammer in the $i$-th end. 
Since the probability of Team $X$ winning the end does not change when they lack the hammer, 
$p_X$ coincides with the probability of Team $X$ winning without the hammer. 
The expected number of points scored by $X$ in the $i$-th end is:
\begin{eqnarray}
E[\mbox{Score i-th end}] = E[Z] \ Pr(X \text{ wins end i}) = \mu\left[q_i \ p_{X,\hammer}+(1-q_i)p_X\right],\label{eq8}    
\end{eqnarray}
where $E[Z] = \mu$ is the expected score for a non-blank end, see equation (\ref{eq7}), and $p_{X,\hammer}$ is the probability of Team $X$ winning with the hammer. We approximate $q_i$ as $\frac{p_Y}{p_X+p_Y}$, the long-term probability of Team~$X$ holding the hammer, see \eqref{eq3}. Multiplying by the 10 ends in a regulation game, and recalling \eqref{eq1}, gives the equilibrium expected score as
\begin{eqnarray}
E[\mbox{Score}]=10\mu\left[\frac{e^{\beta}p_Xp_Y}{(1 - p_X +e^{\beta}p_X)(p_X+p_Y)}+\frac{p^2_X}{p_X+p_Y}\right].\label{eq9}
\end{eqnarray}
To determine match outcome probabilities from estimates of the parameters $\beta$, $\theta$, $p_X$ and $p_Y$, a Monte Carlo simulation of the curling match is performed as follows:
\begin{enumerate}
    \item Simulate the trial of skill to determine the initial Last Stone First End advantage, with Team $X$ winning with probability $\frac{p_X}{p_X + p_Y}$ and Team $Y$ with $\frac{p_Y}{p_X + p_Y}$, proportional to their chances of winning an end.

    \item Suppose, without loss of generality, Team $X$ wins the trial of skill and holds the hammer. There are three scenarios: Team $X$ wins the end with probability $p_{X,\hammer}$ given by \eqref{eq1}, passing the hammer to $Y$; Team $Y$ wins with probability $p_Y$, and $X$ retains the hammer; or, with probability 
    $1 - p_{X,\hammer} -p_Y$ the end is blanked, and $X$ keeps the hammer.
    \item If Team $X$ or Team $Y$ wins the end, the points scored is simulated from the distribution in equation (\ref{eq7}). 
    \item Steps 2-3 are repeated from the perspective of the team with the hammer, until all 10 ends are completed. If the score is tied after 10 ends, step 2 is repeated from the perspective of the team holding the hammer, continuing until one teams wins.
\end{enumerate}
The algorithm for simulating sudden death endings is quite simple. Suppose Team $X$ has the hammer at the 10th end.  Sample from the set $\{0, 1, 2\}$ with replacement using the probabilities from Step 2:
$$
\left(\frac{1+p_Xp_Y-p_X-p_Y-p_Xp_Ye^{\beta}}{1+e^{\beta}p_X-p_X}, \frac{e^{\beta}p_X}{1+e^{\beta}p_X-p_X}, p_Y\right).
$$
The first non-zero value sampled indicates the winning team: a value of 1 means Team $X$ wins, while a value of 2 means Team $Y$ wins. After determining the winner, simulate again using the distribution in equation (\ref{eq7}) to find the number of additional points scored in the winning sudden-death end.

\section{Strategic application}

In this section, we examine a strategic issue in curling explored by Brenzel et al. (2019): When should a team with the hammer opt to blank the end rather than score an additional point, thus surrendering the advantage of holding the hammer? Brenzel et al. (2019) frame this problem in terms of a team's winning probability, denoted as $\winp(x_1, x_2, x_3)$, which depends on three key factors: the current score differential between the two teams ($x_1$), the end number ($x_2$), and whether the team holds the hammer ($x_3$), with $x_3 = 1$ indicating possession of the hammer and $x_3 = 0$ otherwise. According to Brenzel et al. (2019), the team should blank the end if the following condition is met:
\begin{eqnarray}
    \winp(x, e+1, 1)> \winp(x+1, e+1, 0).\label{eq10}
\end{eqnarray}
The inequality in \eqref{eq10} suggests that blanking the end and retaining the hammer is preferable when it enhances the winning probability compared to the alternative of scoring one point and losing possession of the hammer.

The probabilities in equation (\ref{eq10}) can only be determined through Monte Carlo simulation. In contrast, we offer an approximate analytical solution to the blanking problem. This can be readily adjusted for teams of varying strengths, based on the econometric findings presented in Section 5.

Suppose we are currently at end $i$, with Team $X$ holding the hammer and having scored $x$ points, while Team $Y$ has $y$ points. If the current end is blanked, there will be $[10-i-1]$ remaining ends. In the next end, Team $X$ retains the hammer. Consequently, Team $X$ wins the end with probability $p_{X,\hammer}$, while Team $Y$ wins with probability $p_Y$. The remaining $[10-i-2]$ ends are assumed to be in equilibrium, given the lack of detailed information about which team holds the hammer. Based on equation \eqref{eq3}, the equilibrium probability of Team $X$ holding the hammer is given by $q = \frac{p_Y}{p_X + p_Y}$. The expected scores for both teams in this situation are:
\begin{equation*}
\begin{aligned}
 E[\mbox{Team X}] & =  x+ E[Z] \ p_{X,\hammer}  + (10-i-2) \ E[Z] \ [ q \ p_{X,\hammer} + (1-q) p_X] \\  
 & =  x+ \mu  \left[\frac{e^{\beta}p_X}{1 -p_X+e^{\beta}p_X}\right] \\
 & \quad +  (10-i-2)\mu\left[\frac{e^{\beta}p_Xp_Y}{(1 - p_X + e^{\beta}p_X)(p_X+p_Y)}+\frac{p^2_X}{p_X+p_Y}\right]\\
E[\mbox{Team Y}]
 & =  y + E[Z] \ p_Y + (10-i-2) \ E[Z] \ [ (1-q) \ p_{Y,\hammer} + q \ p_Y] \\   
 & = y+ \mu \ p_Y + (10-i-2)\mu\left[\frac{e^{\beta}p_Xp_Y}{(1 - p_Y + e^{\beta}p_Y)(p_X+p_Y)}+\frac{p^2_Y}{p_X+p_Y}\right],
\end{aligned}
\end{equation*}
where $E[Z] = \mu$ is the expected score for a non-blank end, see equation (\ref{eq7}), and $p_{X,\hammer}$ and $p_{Y,\hammer}$ are given in equation \eqref{eq1}.  
In conclusion, assuming that end i is blanked, the expected point difference between the two teams at the end of the match is given by:
\begin{eqnarray}
x-y+\mu\left[\frac{e^{\beta}p_X}{1 - p_X + e^{\beta}p_X}-p_Y\right]+\mbox{Equilibrium adjustment}.\label{eq12}    
\end{eqnarray}
In contrast, if the end is not blanked, the score for Team $X$ increases by 1, and Team $Y$ will have possession of the hammer in the next end. The remaining $[10-i-2]$ ends are assumed to be in equilibrium, as mentioned above. The expected score for both teams in this scenario are:
\begin{equation*} 
\begin{aligned}
 E[\mbox{Team X}] & =    x+1+ \mu \ p_X + (10-i-2)\mu \left[\frac{e^{\beta}p_Xp_Y}{(1 - p_X + e^{\beta}p_X)(p_X+p_Y)}+\frac{p^2_X}{p_X+p_Y}\right] \\
E[\mbox{Team Y}]
 & =  y+ \mu \left[\frac{e^{\beta}p_Y}{1 - p_Y + e^{\beta}p_Y}\right] \\
 & \quad +  (10-i-2)\mu\left[\frac{e^{\beta}p_Xp_Y}{(1 - p_Y + e^{\beta}p_Y)(p_X+p_Y)}+\frac{p^2_Y}{p_X+p_Y}\right].
 \end{aligned}
\end{equation*}
In conclusion, assuming that Team $X$ scores 1 point in end i, the expected point difference between the two teams at the end of the match is given by:
\begin{eqnarray}
x-y+1+\mu\left[p_X-\frac{e^{\beta}p_Y}{1 -p_Y +e^{\beta}p_Y}\right]+\mbox{Equilibrium adjustment}.\label{eq14}    
\end{eqnarray}
By comparing equations (\ref{eq12}) and (\ref{eq14}), it can be concluded that blanking end i is optimal if
\begin{eqnarray}
x-y+\mu\left[\frac{e^{\beta}p_X}{1-p_X+e^{\beta}p_X}-p_Y\right] & {\geq} & x-y+1+\mu\left[p_X-\frac{e^{\beta}p_Y}{1-p_Y+e^{\beta}p_Y}\right],\nonumber\\ \left[\frac{e^{\beta}p_X}{1-p_X+e^{\beta}p_X}+\frac{e^{\beta}p_Y}{1-p_Y+e^{\beta}p_Y}-p_X-p_Y\right]& {\geq} & \frac{1}{\mu}.\label{eq15}  
\end{eqnarray}
Equation (\ref{eq15}) indicates that blanking is optimal only if the advantage both teams gain by holding the hammer exceeds $\frac{1}{\mu}$, where $\mu$ represents the expected value of a scoring end. Consequently, the optimal strategy depends on $p_X$ and $p_Y$ (which reflect the respective strengths of the two teams), $\beta$ (which measures the advantage of holding the hammer), and $\mu$ (the expected margin of victory in a scoring end). Section 5 shows how these parameters can be estimated econometrically.

\section{Econometric application}
In this section we estimate the model presented in Section 3, applying it to historical data from 583 men's international curling matches played from 2019 to 2023. The dataset, originally developed in Austin (2024), is available upon request from the authors. The model in Section 3 was specifically designed for empirical estimation, which is carried out in two distinct phases:
\begin{enumerate}
    \item Estimation of the points-scoring distribution $Z$ for a non-blank ends, see equation (\ref{eq7}).
    \item Estimation of end-winning probabilities, taking into account the two teams involved and which team holds the hammer.  
\end{enumerate}
For the first phase, data from all non-blanked ends is gathered, with the score of the $i$-th end recorded as $x_i$. According to equation \eqref{eq7}, the probability of scoring $x_i$ points in an end is given by $\theta^{x_i}/\sum_{n=1}^8 \theta^n$. Thus, for $N$ samples, the log-likelihood function for $Z$ is expressed as:
\begin{eqnarray}
 l(\theta)= \sum_{i=1}^Nx_i\log(\theta)-N\log(\theta+\theta^2+\theta^3+\theta^4+\theta^5+\theta^6+\theta^7+\theta^8).\label{eq16}   \end{eqnarray}
The parameter $\hat{\theta}$ is estimated by numerical maximisation of $l(\theta)$. When applied to our dataset, this yields an estimate of $\hat{\theta}=0.377$ with an estimated standard error of 0.006. From equation~(\ref{eq7}) the fitted score distribution has mean $\hat\mu = 1.601$. 

 For the second phase, following a similar approach to Fry et al. (2021), the model parameters are estimated using a logistic generalised linear model. Dummy variables for each team are included to account for offensive strength. A positive (negative) coefficient for a team's dummy variable indicates above-average (below-average) offensive skill. Similarly, dummy variables for each opposing team are included to reflect defensive strength. A negative (positive) coefficient for an opponent suggests that the team has above-average (below-average) defensive ability. Additionally, the model includes a dummy variable identifying the team starting the match with the hammer, representing an advantage similar to the home-field effect seen in other sports (Boudreaux et al., 2017; Ehrlich and Potter, 2023).

The above leads to a deceptively complex logistic regression problem to determine the end-winning probabilities for the two teams involved in a match. Here, this complexity is resolved by stepwise selection based on minimisation of the Bayesian Information Criterion (Venables and Ripley, 2002). Results for the final model chosen are shown in Table 1. The coefficient of Hammer is positive and significant reflecting the strategic advantage associated of teams holding the hammer (Brentzel et al., 2019). The coefficient estimate of 2.056 obtained also corresponds to a value of $\beta=2.056$ in equation (\ref{eq8}). The remaining results indicate teams that are unusually effective in either their offensive or defensive play. The team parameters for Scotland and Sweden are positive and significant indicating above average offensive strengths. Relatedly, the team parameters for New Zealand, Finland, Denmark, Korea and Poland are all negative and significant indicating below average offensive strengths. The opponent parameters for Scotland, Sweden, Canada and Italy are all negative and significant indicating above average defensive strengths. Similarly, the opponent parameter for Finland is positive and significant indicating below average defensive qualities. Following a similar approach in Austin (2024) these team-level results are summarised in Table 2 using standard bond-rating terminology to denote quality.

The results in Table 1 are also noteworthy for two further reasons. Firstly, results suggest that curling is primarily a defensive sport. Only Scotland and Sweden have an above average offensive strength. Only, Finland has a below average defensive strength. These observations tally with previous suggestions that sports analytics' models may place a deceptively high value upon good defensive play (McHale et al., 2012). Secondly, the relatively few terms appearing in Table 1 reinforce the highly competitive nature of elite-level curling with many of the top teams seemingly interchangeable. To provide further illustration of regression results in Table 1 an example hand calculation for a match between Sweden and Canada is shown in Table 3. 

\begin{table}[h!]
\begin{center}
\begin{tabular}{|l|l|l|l|l|}
\hline
\textbf{Coefficient}	&	\textbf{Estimate}	&	\textbf{E.S.E.}	&	\textbf{t-value}	&	\textbf{p-value}	\\	\hline
Intercept &      -1.487 &    0.040 & -37.191 & 0.000 \\
\hline
Hammer           & 2.056 &    0.045 &  45.730 & 0.000\\
\hline
Opponent=Scotland & -0.510 &    0.080 &  -6.368 &  0.000\\
\hline
Opponent=Sweden &  -0.491 &    0.080 &   -6.115 & 0.000\\
\hline
Team=Scotland &     0.352 &    0.078 &   4.499 & 0.000\\
\hline
Team=New Zealand &   -1.080 &   0.227 &  -4.750 & 0.000\\
\hline
Team=Finland &      -0.569 &   0.137 &  -4.141 &  0.000\\
\hline
Team=Sweden &       0.246 &    0.078 &   3.150 &  0.002\\
\hline
Opponent=Canada &   -0.364 &    0.098 &  -3.723 &  0.000 \\
\hline
Opponent=Italy  &  -0.264 &    0.079 &  -3.352 & 0.000\\
\hline
Team=Denmark   &   -0.396 &   0.106 & -3.723 & 0.000 \\
\hline
Team=Korea    &    -0.331 &   0.097 & -3.396 & 0.001 \\
\hline
Opponent=Finland &   0.455 &    0.134 &   3.387 & 0.001 \\
\hline
Team=Poland &      -0.818  &  0.257 & -3.185 & 0.001 \\ 
\hline
\end{tabular}
\caption{Generalised linear model results applied to historical international curling matches.}\label{tabreg1}
\end{center}
\end{table}

\begin{table}[h!]
\begin{center}
\begin{tabular}{|l|l|l|}
\hline
\textbf{Team} & \textbf{Rating} & \textbf{Interpretation}\\
\hline
Scotland, Sweden & AAA & Above average attack, \\
 & &above average defence \\

\hline
Canada, Italy & AA+ & Above average defence, \\
& & average attack\\
\hline
China, Czech republic, England, Germany, & AA & Average attack,\\
Japan, Netherlands, Norway, Russia, & & average defence\\
Spain, Switzerland, Turkey, USA & & \\
\hline
Denmark, Korea, New Zealand, Poland & AA--  & Average defence, \\
 &  & below average attack\\
\hline
Finland & A+ & Below average attack,  \\
 & & below average defence\\
\hline
\end{tabular}
\caption{Suggested curling-based interpretation of generalised linear model output.}\label{tabreg2}
\end{center}
\end{table}

\begin{table}[h!]
\begin{center}
\begin{tabular}{|l|}
\hline
Suppose that Sweden play Canada. From Table 2 using Team=Sweden,\\
Opponent=Canada gives\\
$\mbox{logit}(p_{Sweden})=-1.487+0.246-0.164,$\\
$p_{Sweden}=0.167.$\\
Similarly, Table 2 with Team=Canada, Opponent=Sweden gives\\
$\mbox{logit}(p_{Canada})=-1.487-0.491=-1.978,$\\
$p_{Canada}=0.122.$\\
If Sweden have the hammer equation (\ref{eq8}) gives that the\\
expected score is\\
$\mu\left[\frac{e^{\beta}p_{Sweden}}{1+e^{\beta}P_{Sweden}-p_{Sweden}}\right]=1.601\left[\frac{e^{2.056}(0.167)}{1+e^{2.056}(0.167)-0.167}\right]=0.977.$\\
Similarly, the expected score for Canada would be ${\mu}p_{Canada}=1.601\times{0.122}=0.195$.\\
If Canada have the hammer equation (\ref{eq8}) gives the expected score as \\
$\mu\left[\frac{e^{\beta}p_{Canada}}{1+e^{\beta}P_{Canada}-p_{Canada}}\right]=1.601\left[\frac{e^{2.056}(0.122)}{1+e^{2.056}(0.122)-0.122}\right]=0.833.$\\
Similarly, the expected score for Sweden would be ${\mu}p_{Sweden}=1.601{\times}0.167=0.267$.\\
Equation (\ref{eq9}) gives that the equilibrium expected scores would be\\
$E[\mbox{Sweden}]=5.667.$\\
$E[\mbox{Canada}]=5.636.$\\
\hline
\end{tabular}
\caption{Example calculations based on generalised linear model output.}\label{tabreg3}
\end{center}
\end{table}

Based on the Monte Carlo simulation algorithm in Section 3 Table 4 summarises the results of 1,000,000 simulated matches between Sweden and Canada based on parameters estimated from the regression in Table 1\footnote{The related R code is available from the authors upon request.}.

\begin{table}[h!]
\begin{center}
\begin{tabular}{|l|l|l|}
\hline
\textbf{Statistic} & \textbf{Sweden} & \textbf{Canada} \\
\hline
Pr(Win) & 0.668 & 0.332  \\
\hline
Mean score & 10.130 & 7.772 \\
\hline
Median score & 10 & 7 \\
\hline
Variance score & 17.131 & 14.611\\
\hline
Covariance score & -0.243 & -0.243\\
\hline
Inter-quartile range score & 6 & 5 \\
\hline
Skewness score & 0.622 & 0.689 \\
\hline
Kurtosis score & 3.395 & 3.438 \\
\hline
\end{tabular}
\caption{Monte Carlo match simulation results for Sweden v Canada. Results based on 1,000,000 simulations.}\label{tabreg4}
\end{center}
\end{table}

\section{Conclusions and further work}
There has been significant recent interest in sports analytics overall (Baker et al., 2022; Singh et al., 2023; Scarf et al., 2022; Fry et al., 2024), with a specific focus on curling (Brenzel et al., 2019; Lawson and Rave, 2020). By abstracting the game's key characteristics, we develop a Markov model for curling matches. The model's Markovian structure allows us to derive analytical results based on the equilibrium distribution. In particular, we offer an alternative analytical approach to a curling strategy problem previously examined by Brenzel et al. (2019). Additionally, the Markovian framework supports analyses through Monte Carlo simulation.

The model developed in this paper consists of two independent components. The first component is the distribution of points in scoring ends, and the second is the end-winning probabilities, which depend on the respective skills of the two teams and their possession of the hammer. This structure enhances the model's tractability and facilitates econometric estimation. A systematic approach to the distribution of points in scoring ends is provided using the maximum entropy method (Visser, 2013). Estimation can be conducted through numerical maximum likelihood techniques. The second component can be addressed using generalised linear modelling of historical results. The findings allow us to quantify the advantage of holding the hammer, and offer new methods for ranking and comparing teams based on their offensive and defensive strengths.

The implications of our paper are as follows. The econometric results allow us to quantify the advantage associated with holding the hammer, and provide new methods for ranking and comparing teams based on their offensive and defensive strengths. Our findings suggest that curling is a highly competitive yet primarily defensive sport. This aligns with previous research indicating that sports analytics often reveals overlooked aspects of strong defensive play (McHale et al., 2012). Our results add to an appreciation of curling, given the game's strategic complexity (Brenzel et al., 2019) and its intensely competitive nature. Notably, half of the teams analysed appear largely interchangeable, with only minor differences observed among elite-level teams. Additionally, sports analytics can be highly relevant in educational contexts (Wooten and White, 2021). As this paper demonstrates, the econometric analysis of historical sports data can yield challenging yet insightful examples of regression and generalised linear modelling. These examples are valuable both computationally and in terms of their practical interpretation. Some teaching examples on a related theme can be found in Fry and Burke (2022). Our study also underscores the significance of contemporary applications of sports analytics, including Markov modelling.

Interest in sports analytics continues to thrive (Ehrlich et al., 2023; Fry et al., 2024; Potter and Ehrlich, 2022), with a particular focus on Markov modelling (Brenzel et al., 2019; Kostuk et al., 2001). Future research will aim to enhance the analytical framework of the model to achieve a closer alignment between analytical and simulation results. Aspects of the Markov model developed in this paper may hold independent theoretical significance. Given the deceptive complexity of curling, further analysis of curling strategies, whether through analytical methods or simulation, remains a compelling area for investigation.

\end{document}